\documentclass[aps,prd,10pt,twocolumn,showpacs,preprintnumbers,amsmath,amssymb,floatfix,nofootinbib]{revtex4-1}
\usepackage[hyperfootnotes=true]{hyperref} 
\pdfoutput=1

\usepackage{natbib} 
\usepackage{amsmath}
\usepackage{epsfig}
\usepackage{array}
\usepackage{color}

\begin{document}


\title{ Cosmological Perturbation Theory as a Tool for Estimating \\ 
Box-Scale Effects in $N$-body Simulations     }
        
\author{ Chris Orban$^{1,2}$}\email{orban@physics.osu.edu}

\affiliation{
\vspace{0.2cm}
(1) Center for Cosmology and Astro-Particle Physics, The Ohio State University, 191 W Woodruff Ave, Columbus, OH 43210 \\
(2) Department of Physics, The Ohio State University, 191 W Woodruff Ave, Columbus, OH 43210 
}


\begin{abstract}
In performing cosmological N-body simulations, it is widely appreciated that the growth of structure on the largest scales within a simulation box will be inhibited by the finite size of the simulation volume. Following ideas set forth in \citet{Seto1999}, this paper shows that standard (a.k.a. 1-loop) cosmological perturbation theory (SPT; \cite{Vishniac1983}) can be used to predict, in an approximate way, the deleterious effect of the box scale on the power spectrum of density fluctuations in simulation volumes. Alternatively, this approach can be used to quickly estimate {\it post facto} the effect of the box scale on power spectrum results from existing simulations. In this way SPT can help determine whether larger box sizes or other more-sophisticated methods are needed to achieve a particular level of precision for a given application (e.g. simulations to measure the non-linear evolution of baryon acoustic oscillations). I focus on SPT in this note and show that its predictions differ only by about a factor of two \emph{or less} from the measured suppression inferred from both powerlaw and $\Lambda$CDM $N$-body simulations. It should be possible to improve the accuracy of these predictions through using more-sophisticated perturbation theory models. An appendix compares power spectrum measurements from the powerlaw simulations at outputs where box-scale effects are minimal to perturbation theory models and previously-published fitting functions. These power spectrum measurements are included with this paper to aid efforts to develop new perturbation theory models.
\end{abstract}


\maketitle

\section{Introduction}
  \label{sec:intro}

Over many years a number of tools have been developed to simulate and understand the inherently non-linear process of structure growth in the universe. Perturbation theory models have become substantially more accurate 
(e.g. \cite{Crocce_scoccimarro2006,MatarresePietroni2007,Pietroni2008,Matsubara2008,Hiramatsu_Tauruya2009,Crocce_etal2012} and see \cite{Carlson_etal09} for comparisons),
and cosmologists routinely use a variety of sophisticated N-body simulation methods to efficiently make predictions for the clustering of the dark matter distribution \cite{Schneider_etal2011,Angulo_White2010,Tassev_Zaldarriaga2012,Tassev_etal2013,Jenkins2013}. Many of these tools are central to extracting cosmological information from current or near future surveys.

This paper extends this toolkit with a method to estimate the \emph{inaccuracy} of $N$-body simulations on scales approaching the box size. Originally proposed by \citet{Seto1999} but not validated with simulation results, it will be shown that cosmological perturbation theory can provide usefully accurate model of the suppression of the growth of structure as a result of the finiteness of the simulation volume\footnote{Box-size effects were also studied in detail in \cite{Takahashi_etal2008}. Although that paper also uses perturbation theory to assess and predict the effect of the finite scale of the simulation box, they focused on the \emph{box-to-box variance} in the mean power spectrum  whereas the present work considers how the mean power spectrum can be \emph{biased} from the finiteness of the box.}. This insight will help assess or, in planning, predict the accuracy of cosmological N-body simulations that are used to interpret results from cosmological surveys \cite{Smith_etal2014}.

Although this paper focuses on the suppression of the power spectrum, the method outlined here is relevant to box-scale effects on other statistics, such as the two-point correlation function or the halo mass function, through their relation to the power spectrum. Importantly, the method does not require knowledge of the initial phases of the Fourier modes in the simulation which means that it can be applied quite generally even for simulations where the initial conditions have been lost or are unavailable. Note well that this phase-independence implies that the perturbation theory models are here used to estimate the suppression of the \emph{mean} power spectrum in \emph{ensembles} of simulations. This estimate can still be relevant for individual simulations as a rule-of-thumb for how large the box-scale effects are likely to be.

This paper demonstrates this approach using N-body simulations with power-law initial conditions performed by the author and comparing to previously-published $\Lambda$CDM simulation results from \cite{Heitmann_etal2010}. The reason for the focus on powerlaw simulations is two-fold: (1) \citet{Seto1999} investigated powerlaw initial conditions in creating SPT predictions for the box-scale suppression, and (2) the self-similarity of the initial conditions can be used to assess the accuracy of the non-linear power spectrum on scales approaching the box without the need for performing a significantly more ambitious simulation ensemble with a larger box and significantly more particles. To explain more of this latter point, the box-scale results at a particular output, for example, can be compared to appropriately-scaled results from an earlier output when box scale effects should be less of a concern. If the two results agree then the finiteness of the simulation volume is understood to have had negligible effect. In this paper self-similarity is used in a more sophisticated way than just described but this is the essential idea.

It should also be said that the simplicity of powerlaw initial conditions is very convenient from a theoretical standpoint. Analytic solutions exist for powerlaw initial conditions for the standard (a.k.a. 1-loop) perturbation theory (SPT) scheme \cite{Vishniac1983,Makino_etal1992,Scoccimarro_Frieman1996,Scoccimarro1997,Pajer_Zaldarriaga2013} and the self-similarity of the initial conditions can ameliorate the difficulty of calculating the other non-linear statistics as well (e.g. \cite{Bernardeau_etal2002}). Thus power-law initial conditions continue to find relevance in modern research, from developing new perturbation theory models \cite{Pajer_Zaldarriaga2013,Mercolli_Pajer2013} or testing cosmological N-body simulation methods \cite{Orban2013}, or examining the non-linear physics of baryon acoustic oscillations (BAO) \cite{Orban_Weinberg2011} among other examples. In the present case powerlaw simulations (specifically $n = -1.5$ \& 2) provide a clean and convenient way of assessing the usefulness of perturbation theory at the box scale. 

As already mentioned, SPT estimates for the box-scale suppression are also compared, in Sec.~\ref{sec:lcdm}, to $N$-body simulation results presented in \cite{Heitmann_etal2010}. This comparison is less rigorous than the comparison to powerlaw simulation results. Rules-of-thumb for the box-scale suppression presented in other studies are discussed and an SPT-inspired rule-of-thumb is given. Projections are made for the magnitude of box-scale effects at one-fifth the box scale in $\Lambda$CDM simulations with various box sizes and redshifts which can be validated in future studies.

\section{Powerlaw Simulations}
\label{sec:powlaw}

This study presents two ensembles of simulations using the publicly-available Gadget2 code (Springel 2005) to integrate particle trajectories from the initial conditions. Gadget2 is a well-tested and efficient Tree-PM code that compares well to other N-body codes in use \cite{Heitmann_etal2010}. Gadget2 was used to perform a number of dark-matter-only simulations using $N=512^3$ particles\footnote{n.b. the box-scale focus of this study did not require simulations with significantly larger numbers of particles} and a $1024^3$ PM grid. No modifications to the code were required to perform these simulations. As long as the dark matter density is set to the critical density (i.e. $\Omega_m = 1$) the comoving density field should evolve in a self-similar way. 

The publicly-available 2LPT code \cite{Crocce_etal06} was used to generate power-law initial conditions. This code computes the initial particle displacements and velocities using the Zeldovich approximation \cite{Zeldovich1970} and second-order corrections from Lagrangian perturbation theory. The inclusion of these second-order corrections has been found to significantly improve the realism of the initial conditions and minimize numerical transients at the beginning of simulations. The initial epoch of the simulations was determined setting the dimensionless power at the particle Nyquist frequency, $\Delta_{ic}^2(k_{Ny,p})$ where $k_{Ny,p} = \pi N^{1/3} / L_{\rm box}$, to be near or below 0.001. 

2LPT was used with an initial $P_L (k) = A \, k^n$ spectrum where $P_L (k)$ is the linear power spectrum, $k$ is the wavenumber of the density perturbation, $A$ is a factor that scales as the square of the linear growth function and $n$ is the power law. This study focuses on $n = -1.5$ and $n = -2$. Note that the effective slope of the $\Lambda$CDM linear power spectrum is similar to $n = -1.5$ on BAO scales \cite[][Fig.~2]{Orban_Weinberg2011}. The $n = -2$ spectrum corresponds to the effective slope of the $\Lambda$CDM spectrum on somewhat smaller scales. 

For each powerlaw, 25 simulations were performed for the purpose of accumulating statistics of the evolved dark matter density field. The $n = -2$ simulations presented here are similar to Widrow et al. \cite{Widrow_etal09}, which present an $n = -2$ simulation with a significantly higher particle count ($N = 1024^3$). However that study presents only one realization of this density field. As a result the error bars on their self-similar fits to the non-linear power spectrum results -- errors which were not shown or estimated -- can be quite large on quasi-linear scales, i.e., the scales that they were not particularly interested in. The two simulation ensembles presented here are designed to investigate these quasi-linear scales at high precision. In this sense these are the most precise measurements of the self-similar behavior of $n = -1.5$ and $n = -2$ powerlaws in the literature. 

Developers of new perturbation theory schemes may find this data set useful as a cross-check of the accuracy of perturbation theory methods, accordingly these data are included with this paper. As an example, Appendix~\ref{ap:sf96} compares the measured power spectra from the simulations to the SPT model \cite{Vishniac1983}. Appendix~\ref{ap:sf96} shows that SPT typically predicts the non-linear evolution of the power spectrum to within a few percent on quasi-linear scales for both $n = -1.5$ and $n = -2$ cases. This result is significant for the next section where SPT predictions with and without a truncated linear power spectrum will be compared in order to estimate the box-scale suppression of the non-linear power spectrum.

\section{Suppression of Structure Growth from Finite Volume Effects}
\label{sec:ptsuppress}

Seto \cite{Seto1999} provides arguments that the {\it integral expressions}
for SPT can be used with a truncated initial power spectrum to capture the effect of the box scale on the growth of structure in cosmological N-body simulations. Formally, then, if 
\begin{equation}
P_{L,\rm{trunc}}(k) = \left\{ 
\begin{array}{l l}
  0 & \quad  k < k_{\rm{box}} \\
  A k^n & \quad k_{\rm{box}} < k < k_c \\ 
  0  & \quad k > k_c \\ \end{array} \right. 
\end{equation}
represents the initial power spectrum up to some cutoff, $k_c$, then the predicted non-linear power spectrum for $k \gtrsim k_{\rm{box}}$ from SPT is 
\begin{equation}
P_{\rm{SPT,trunc}}(k) = P_L(k) + P_{2,\rm{trunc}}(k) \label{eq:spttrunc}
\end{equation}
where
\begin{equation}
P_{2,\rm{trunc}}(k) = P_{22,\rm{trunc}}(k) + P_{13,\rm{trunc}}(k)
\end{equation}
and
\begin{widetext}
\begin{equation}
\displaystyle P_{22,\rm{trunc}}(k) =  \frac{k^3}{98 \left(2 \pi\right)^2}   \int_0^\infty dq \, P_{L,\rm{trunc}} \left(q\right) \int_{-1}^1 {dx} \,  P_{L,\rm{trunc}}\left[\left(k^2+q^2 -2 k q x\right)^{1/2}\right]\frac{\left(3 q +7 k x-10 q x^2\right)^2}{\left(k^2+q^2-2 k q x\right)^2} \label{eq:P22}
\end{equation}
\begin{equation}
\displaystyle P_{13,\rm{trunc}}(k)= \frac{k^3 P_{L,\rm{trunc}} (k)}{252 \left(2 \pi\right)^2}\int_0^\infty dq P_{L,\rm{trunc}} (q) \Big[\frac{12k^2}{q^2}-158+100 \left( \frac{q}{k} \right)^2   - 42 \left( \frac{q}{k} \right)^4 + \frac{3 k^3}{q^3}\left[\left(\frac{q}{k}\right)^2-1\right]^3 \left(7 \left(\frac{q}{k}\right)^2 +2\right)\ln\bigg|\frac{k+q}{k-q}\bigg|\Big]. \label{eq:P13}
\end{equation}
\end{widetext}
\citet{Seto1999} shows the ratio $P_{2,\rm{trunc}}(k) / P_2(k)$ (i.e. truncated over no truncation) versus $k / k_{\rm{box}}$ for $n = 1, 0 , -1$ and $-2$ as an expectation for how much the box scale can change the non-linear 
growth. However \citet{Seto1999} does not actually validate these predictions with $N$-body simulations.

In this paper the box scale suppression will be estimated as the ratio $P_{\rm SPT,trunc}(k) / P_{\rm SPT} (k)$ instead of $P_{\rm SPT, trunc}(k) / P_{\rm true}(k)$ where $P_{\rm true} (k)$ represents the true self-similar power-spectrum, usually determined by high-resolution $N$-body simulation results\footnote{Although we use the label ``true'' to describe results from high-resolution $N$-body simulations, this is not to imply that high-resolution results are entirely without inaccuracies. With this in mind Appendix~\ref{ap:sf96} presents self-similar scaling tests that show (right-hand column of Fig.~\ref{fig:sf96}) that the fitting functions for the non-linear power spectra from high-resolution simulations of initially $n = -1.5$ and $-2$ powerlaws (i.e. the ``true'' non-linear power spectrum models assumed here) are accurate at the few-percent level or better. Appendix~\ref{ap:sf96} arrives at this conclusion in a conservative way using only the earliest few simulation outputs and clustering measurements on spatial scales many times smaller than the box size ($k > 5 \, k_{\rm box}$). This few-percent-or-better accuracy is sufficient for the purposes of the next section where simulations are evolved to such high clustering levels that the box-scale suppression of the non-linear power spectrum reaches 10-25\% at one-fifth of the box scale.}. In practice, it is better and more convenient to compute the ratio $P_{\rm SPT,trunc}(k) / P_{\rm SPT}(k)$ (which is equivalent to  $\Delta_{\rm SPT,trunc}^2 (k) / \Delta_{\rm SPT}^2 (k)$, an expression that is used later\footnote{This is because $P (k)$ is related to the dimensionless power spectrum in the usual way: $\Delta^2 (k) = k^3 P (k) / 2 \pi^2$.}) because on very weakly non-linear scales or early-enough epochs both $P_{\rm SPT,trunc}(k)$ and $P_{\rm SPT}(k)$ converge to the linear power spectrum. Thus the expression always asymptotes to one indicating no box-scale suppression as would be expected. This approach also sidesteps the few-percent inaccuracies of SPT on quasi-linear scales (c.f. Appendix~\ref{ap:sf96}), which would otherwise contaminate high-precision estimates of box scale effects, e.g., for the non-linear evolution of the BAO feature. 

The analytic results from Appendix B of \cite{Scoccimarro_Frieman1996}, \cite{Scoccimarro1997} and in the more recent study by \cite{Pajer_Zaldarriaga2013} are quite helpful in eliminating the need for numerical integration to obtain $\Delta_{\rm SPT}^2(k)$, which is the SPT dimensionless power spectrum \emph{without} truncation. In the studies just mentioned the dimensionless power spectrum is of the form
\begin{equation}
\Delta_{\rm SPT}^2(k) = \Delta_L^2(k) + \Delta_2^2(k) = \Delta_L^2(k) \left( 1 + \alpha_n \Delta_L^2(k) \right) \label{eq:spt}
\end{equation}
where $\alpha_n$ is a constant and
\begin{equation}
\Delta_L^2(k) = \left(\frac{k} { k_{\rm{nl}}}\right)^{n+3} \label{eq:SF_eq1}
\end{equation}
thus
\begin{equation}
\Delta_2^2(k) = \alpha_n (\Delta_L^2(k) )^2 = \alpha_n \left( \frac{k} {k_{\rm{nl}}} \right)^{2(n+3)}. \label{eq:SF_eq2}
\end{equation}
In developing their Effective Field Theory (EFT) model, Pajer \& Zaldarriaga \cite{Pajer_Zaldarriaga2013}
realized that the SPT expression for $\alpha_n$ in \cite{Scoccimarro_Frieman1996} and \cite{Scoccimarro1997} is incorrect. Pajer \& Zaldarriaga report $\alpha_{-1.5} = 0.239$ and $\alpha_{-2} = 1.38$ as the correct values and supply a mathematica notebook to verify the result. For the present study, these values were also confirmed using numerical integration. Appendix~\ref{ap:sf96} shows that these numbers for $\alpha_n$ agree better with simulation results. 

Using Eqs.~\ref{eq:spt}-\ref{eq:SF_eq2} and Eqs.~\ref{eq:spttrunc}-\ref{eq:P13},
\begin{eqnarray}
\frac{\Delta_{\rm SPT,trunc}^2(k)}{\Delta_{\rm SPT}^2(k)} = & \displaystyle \frac{\Delta_L^2(k) + \Delta_{2,\rm{trunc}}^2(k)}{\Delta_L^2(k) + \Delta_2^2(k)} \nonumber \\
= & \displaystyle   \frac{ 1  + \Delta_{2,\rm{trunc}}^2(k)/\Delta_L^2(k)}{ 1 + \Delta_2^2(k)/\Delta_L^2(k)} \nonumber \\
= & \displaystyle   \frac{1 + \alpha_n (k / k_{\rm nl})^{n+3} \left[ \Delta_{2,\rm{trunc}}^2(k)/\Delta_2^2(k) \right] }{1 + \alpha_n  (k / k_{\rm{nl}})^{n+3} }. \label{eq:seto}
\end{eqnarray}
The term, $\Delta_{2,\rm{trunc}}^2(k)/\Delta_2^2(k)$, is independent of epoch and
is identical to the $P_{2,\rm trunc}(k) / P_2 (k)$ quantity plotted in Fig. 3 of \citet{Seto1999}. $\Delta_{2,\rm trunc}^2(k)$ is evaluated numerically using 
the publicly-available ``copter'' code developed by Jordan Carlson \cite{Carlson_etal09}. For best results, empirically it was found that the high-$k$ cutoff ($k_c$) in the numerical integration should be taken orders of magnitude larger than the particle Nyquist frequency would suggest. This is analogous to computing box-scale effects for simulations with a finite box but with near-infinite spatial resolution. 

\section{Powerlaw Results}
\label{sec:results}

\begin{figure*}
\centerline{\epsfig{file=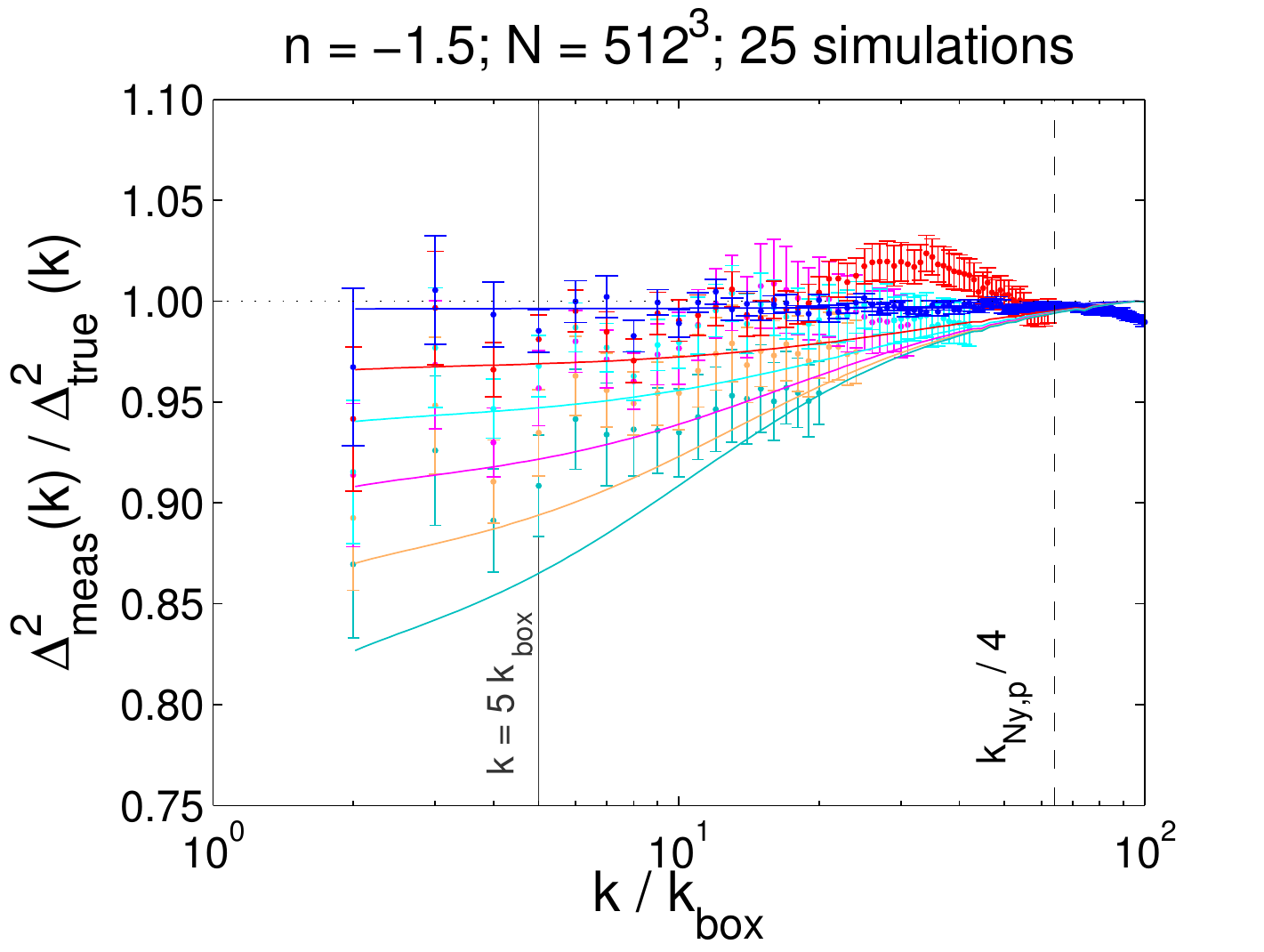,width=3.1in}\epsfig{file=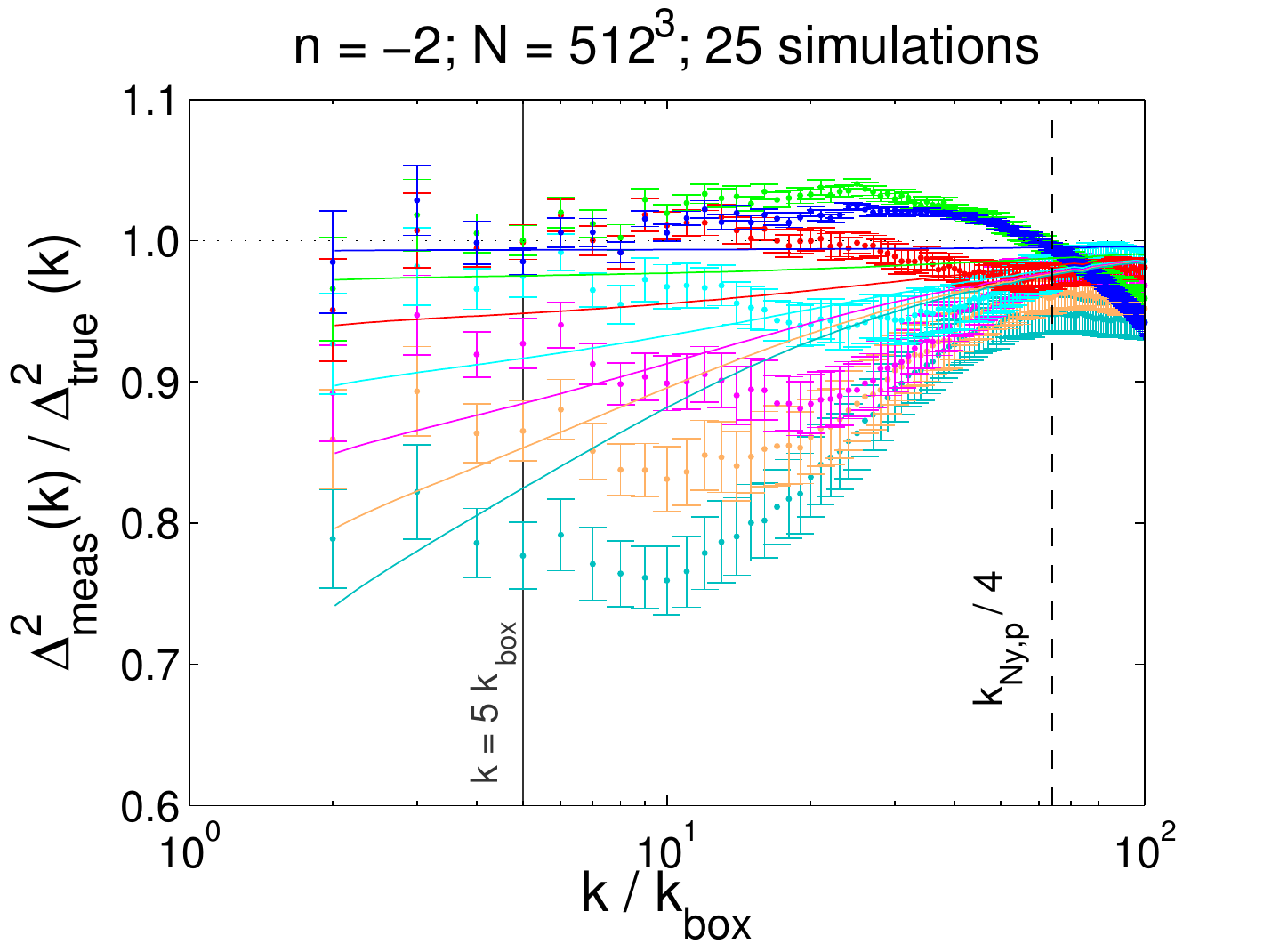,width=3.1in}\epsfig{file=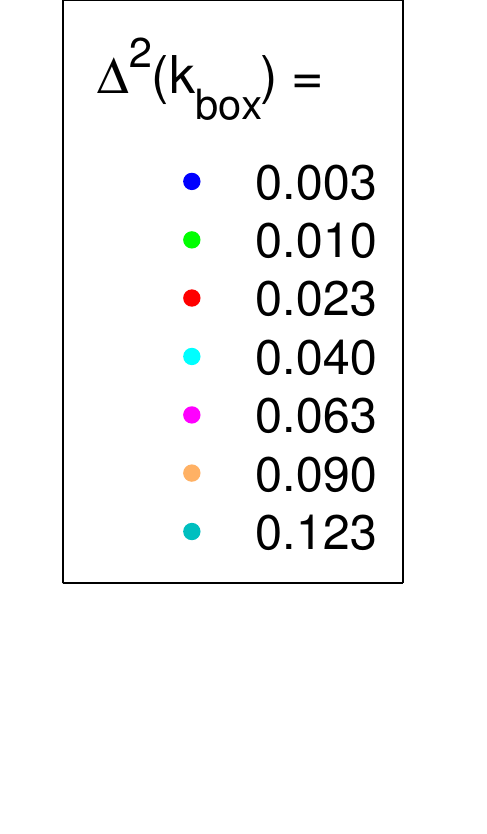,width=1in}}
\caption{Comparing the measured power spectrum from ensembles of 25 N-body simulations ($N=512^3$) to the true self-similar power spectrum, $\Delta_{\rm true}^2(k)$, relative to the box scale of the simulation volume ($k_{\rm box} = 2 \pi / L_{\rm box}$). The left panel shows $n = -1.5$ powerlaw results while the right panel shows $n = -2$ results. Colors correspond to simulation outputs according to the legend on the far right. Solid lines show the predicted suppression of the mean power spectrum according to the \citet{Seto1999}-inspired estimate in Eq.~\ref{eq:seto}. Error bars show the 1$\sigma$ error on the mean measured from the 25 realizations.
}\label{fig:ptmiss}  
\end{figure*}

Fig.~\ref{fig:ptmiss} compares the predicted box-scale suppression of the mean power spectrum from Eq.~\ref{eq:seto} (lines of various colors) to the $N$-body simulation results (colored points with error bars) for both $n = -1.5$ (left panel) and $n = -2$ (right panel). Colors correspond to a given level of box scale clustering, $\Delta_L^2 (k_{\rm box})$, as indicated in the legend to the right of the $n = -2$ results. The simulations are evolved until $\Delta_L^2 (k_{\rm box}) = 0.123$.

\begin{figure*}
\centerline{\epsfig{file=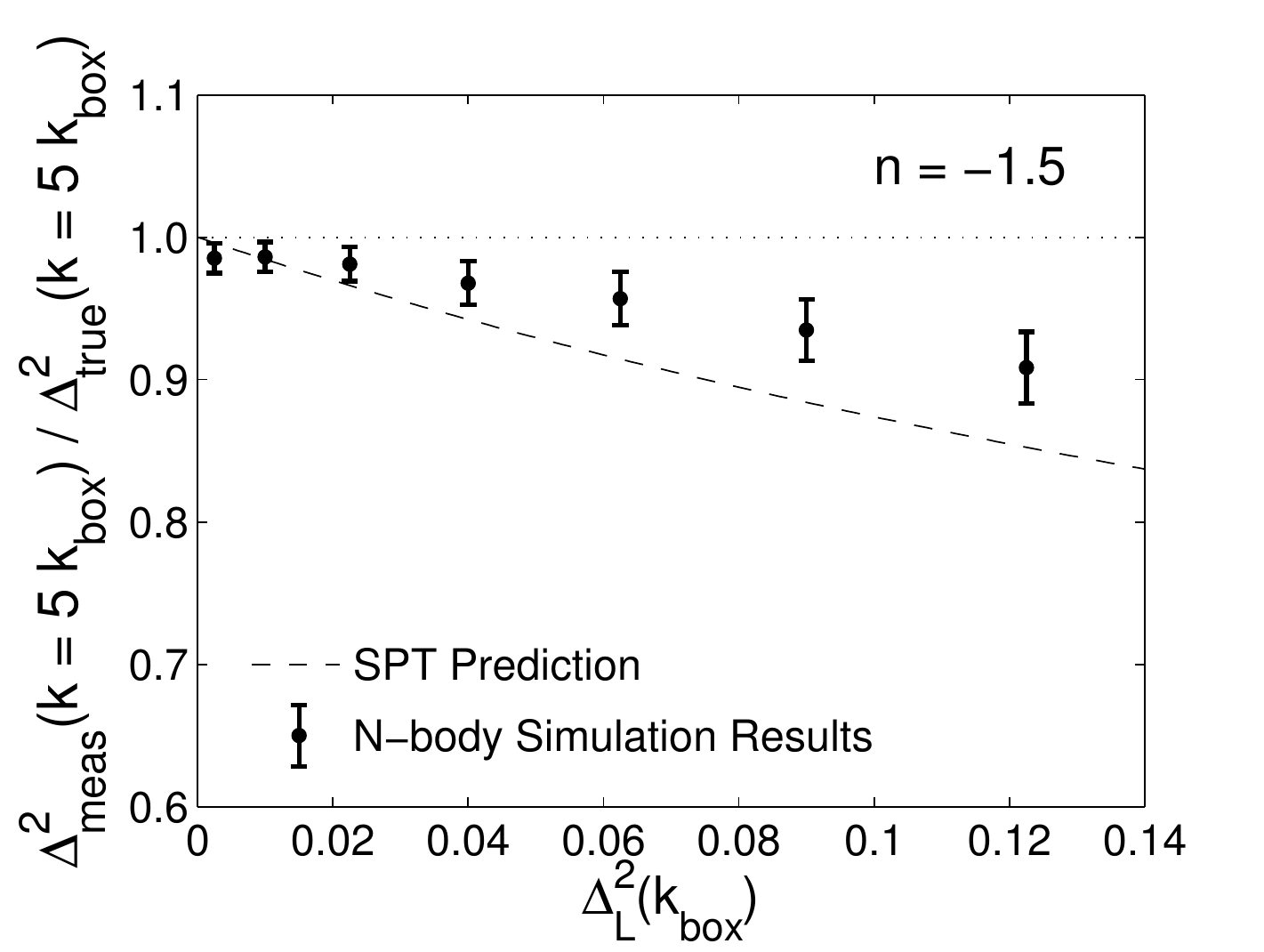,width=3.1in}\epsfig{file=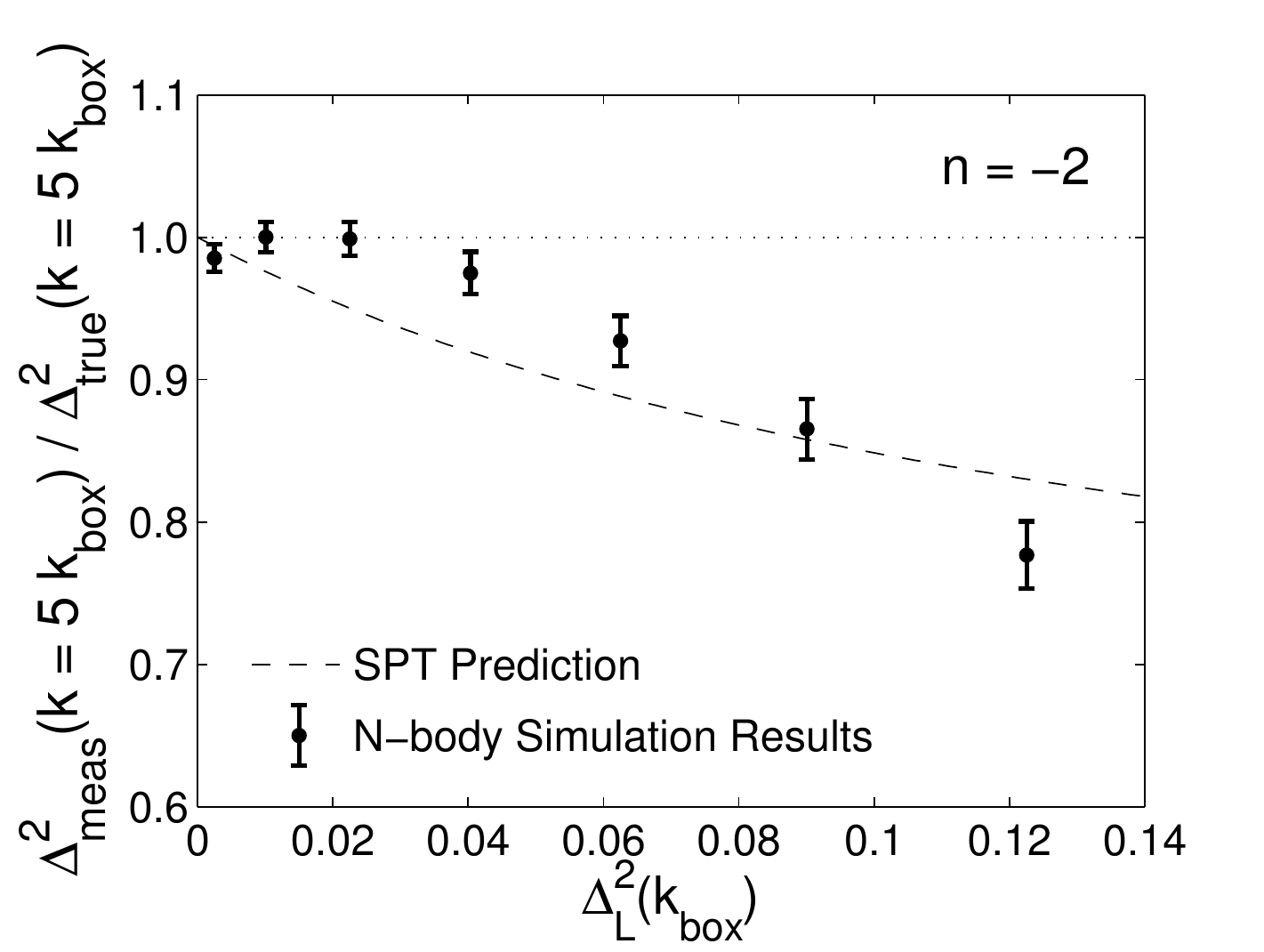,width=3.1in}}
\caption{ Comparing the estimated (Eq.~\ref{eq:seto}, in black dashed lines) and measured (solid black points with error bars) suppression of the mean power spectrum at one fifth of the box scale ($k = 5 \, k_{\rm box}$). The x-axis shows the level of dimensionless (linear) clustering power on the scale of the simulation box, $\Delta_L^2(k_{\rm box})$, which becomes larger as the cosmological density field is evolved. }\label{fig:Deltasq}
\end{figure*}

Although the measured 1$\sigma$ error bars on the mean power spectrum from 25 simulations are still relatively large in some cases, Fig.~\ref{fig:ptmiss} confirms a qualitative resemblance between the measurements from simulation and the predictions of truncated SPT (Eq.~\ref{eq:seto}). The $n = -2$ case is clearer to interpret than the $n = -1.5$ results in part because $\Delta_{\rm true}^2 (k)$ comes from the high-resolution simulation from Widrow et al. \cite{Widrow_etal09} whereas $\Delta_{\rm true}^2(k)$ for the $n = -1.5$ case (which was not considered in Widrow et al. \cite{Widrow_etal09}) comes from the fit to comparatively lower resolution simulations ($N = 512^3$) presented in Appendix A of Orban \& Weinberg \cite{Orban_Weinberg2011}. At early outputs ($\Delta_L^2(k_{\rm box}) \lesssim 0.023$) where one expects the measured power spectrum to match the self-similar power spectrum to good accuracy one does find that the simulations match $\Delta_{\rm true}^2(k)$ to within a few percent (which is similar to the accuracy of Widrow's determination of $\Delta_{\rm true}^2(k)$ for $n = -2$) and even at very high $k / k_{\rm box}$ where finite particle effects could cause deviations from Widrow's result. Both panels show one fourth the particle Nyquist frequency with a vertical dashed black line. To the right of this line one does not expect good agreement between these simulations and $\Delta_{\rm true}^2(k)$.

Towards the end of the simulations the clustering level on the scale of the box becomes quite large, $\Delta_L^2 (k_{\rm box})~\sim~0.1$, and both the $n = -1.5$ and $n = -2$ results at low $k / k_{\rm box}$ show a clear trend with $\Delta_L^2(k_{\rm box})$ with $\sim10$\% level suppression of the mean power spectrum in the $n = -1.5$ results and $\sim20$\% level suppression in the $n = -2$ results. From the perspective of Smith et al. \cite{Smith_etal2003} who emphasize that the important quantity is the missing \emph{variance} from the simulations,
\begin{equation}
\sigma_{\rm miss}^2 \approx \int_0^{k_{\rm box}} \Delta^2 (k) \, \frac{dk} { k} \approx \frac{\Delta_L ^2(k_{\rm box})} { n + 3}, \label{eq:sigmamiss}
\end{equation}
where $\sigma_{\rm miss}$ is this missing variance, it is natural that the $n = -2$ results show significantly larger suppression than $n = -1.5$. According to their precise expression for $\sigma_{\rm miss}$, for the same $\Delta_L^2(k_{\rm box})$, $\sigma_{\rm miss}$ is about twice as large for $n = -2$ as for $n = -1.5$. This is qualitatively consistent with the indication from Fig.~\ref{fig:ptmiss} that the suppression is about twice as severe in the $n = -2$ simulations. This conclusion is very interesting for the goal of understanding box-scale effects, however note that Eq.~\ref{eq:sigmamiss} is not shown on Figs.~\ref{fig:ptmiss} \& \ref{fig:Deltasq} because Smith et al. \cite{Smith_etal2003} do not connect $\sigma_{\rm miss}$ with a method to predict the detailed shape of the suppressed power spectrum. In fact, Eq.~\ref{eq:sigmamiss} can more easily be interpreted as the suppression of the \emph{two-point correlation function}, i.e.,
\begin{eqnarray}
\displaystyle \xi_{\rm meas}(r) = \, \, \, \, & \displaystyle \int_{2 \pi / L_{\rm box}}^\infty \Delta^2(k) \frac{\sin (kr)}{kr}  \frac{dk}{k} \nonumber \\
\, \, \, \, \, \, \approx \, \, \, \, & \displaystyle \xi_{\rm true}(r) - \int_{0}^{k_{\rm box}} \Delta^2(k) \frac{dk}{k} \nonumber \\
 = \, \, \, \,  & \displaystyle \xi_{\rm true}(r) - \sigma_{\rm miss}^2 
\end{eqnarray}
where  $\xi_{\rm meas}(r)$ and $\xi_{\rm true}(r)$ are the measured and ``true'' correlation functions for separations, $r$, and the last step assumes $k_{\rm box}$ is small. This result straightforwardly implies that the correlation function is smaller than its true value by $\sigma_{\rm miss}^2$ \cite{Sirko2005}. The relation between $\sigma_{\rm miss}$ and the suppression of the \emph{power spectrum} is less clear. 

Fig.~\ref{fig:Deltasq} quantitatively compares the SPT prediction for the suppression to the simulation results at $k = 5 \, k_{\rm box}$. For reference, $k = 5 \, k_{\rm box}$ is indicated in Fig.~\ref{fig:ptmiss} with a vertical solid black line. Fig.~\ref{fig:Deltasq} shows that the model is capable of making factor-of-two-accurate or better estimates for the suppression of the mean power spectrum. Given that the SPT \emph{without} box-scale truncation compares favorably to the simulation results from the first few outputs (Appendix~\ref{ap:sf96}) it is presently unclear why the agreement in Figs.~\ref{fig:ptmiss} \& \ref{fig:Deltasq} is not better than it is. The early outputs for the $n = -2$ case are particularly concerning since the power spectrum measurements indicate minimal suppression while the model predicts $\sim 5$\% suppression. Also, the model predicts $\sim 15$\% suppression for both powerlaws, whereas the $\sigma_{\rm miss}$ rule-of-thumb (at the same $\Delta_L^2 (k_{\rm box})$) would predict more suppression for $n = -2$ and less suppression for $n = -1.5$. A deeper consideration of these issues is deferred to future work.

\section{Application to $\Lambda$CDM }
\label{sec:lcdm}

\begin{figure}
\epsfig{file=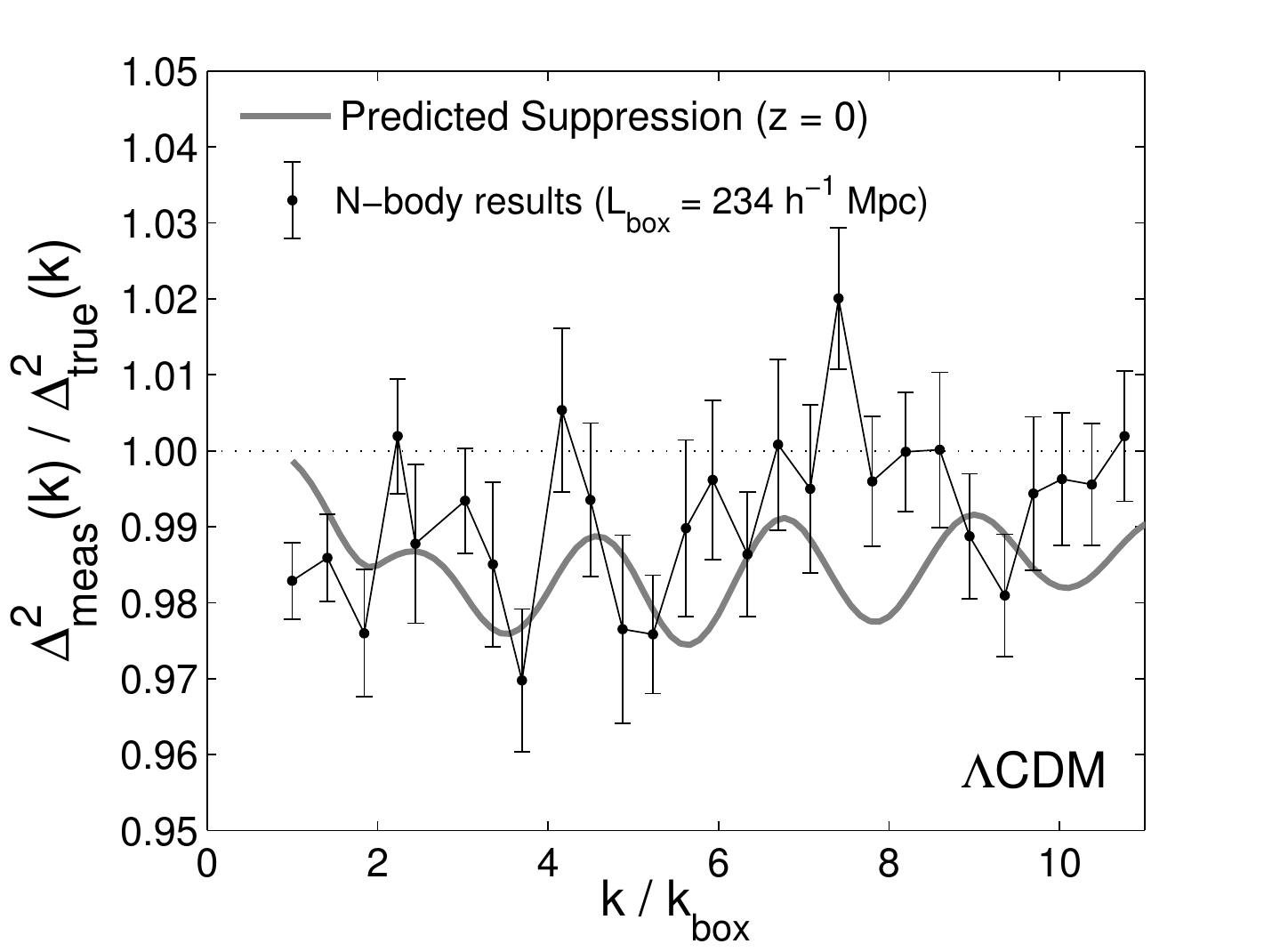,width=3.5in}
\vspace{-0.75cm}
\caption{A comparison between the SPT-estimated box-scale suppression of the power spectrum for a $\Lambda$CDM cosmology (gray line) and the measured box scale suppression inferred entirely from $N$-body simulation results presented in \cite{Heitmann_etal2010} (black line with error bars); specifically Fig.~6 from that study. All results are for $z = 0$. Here $\Delta_{\rm true}^2 (k)$ (i.e. the ``true'' non-linear power spectrum) comes from simulations with $L_{\rm box} = 2000 h^{-1}$Mpc. This box scale is sufficiently large that the suppression of the non-linear power spectrum is negligible on the scale shown. These results are compared to non-linear power spectrum measurements from simulations with a much smaller box scale ($L_{\rm box} = 234 h^{-1}$Mpc). These measurements are referred to as $\Delta_{\rm meas}^2(k)$. Plotting the ratio of $\Delta_{\rm meas}^2(k)$ and $\Delta_{\rm true}^2 (k)$ thus reveals the box-scale suppression of the power spectrum in the $L_{\rm box} = 234 h^{-1}$Mpc simulations.} \label{fig:coyote}
\end{figure}

\begin{figure*}
\centerline{\epsfig{file=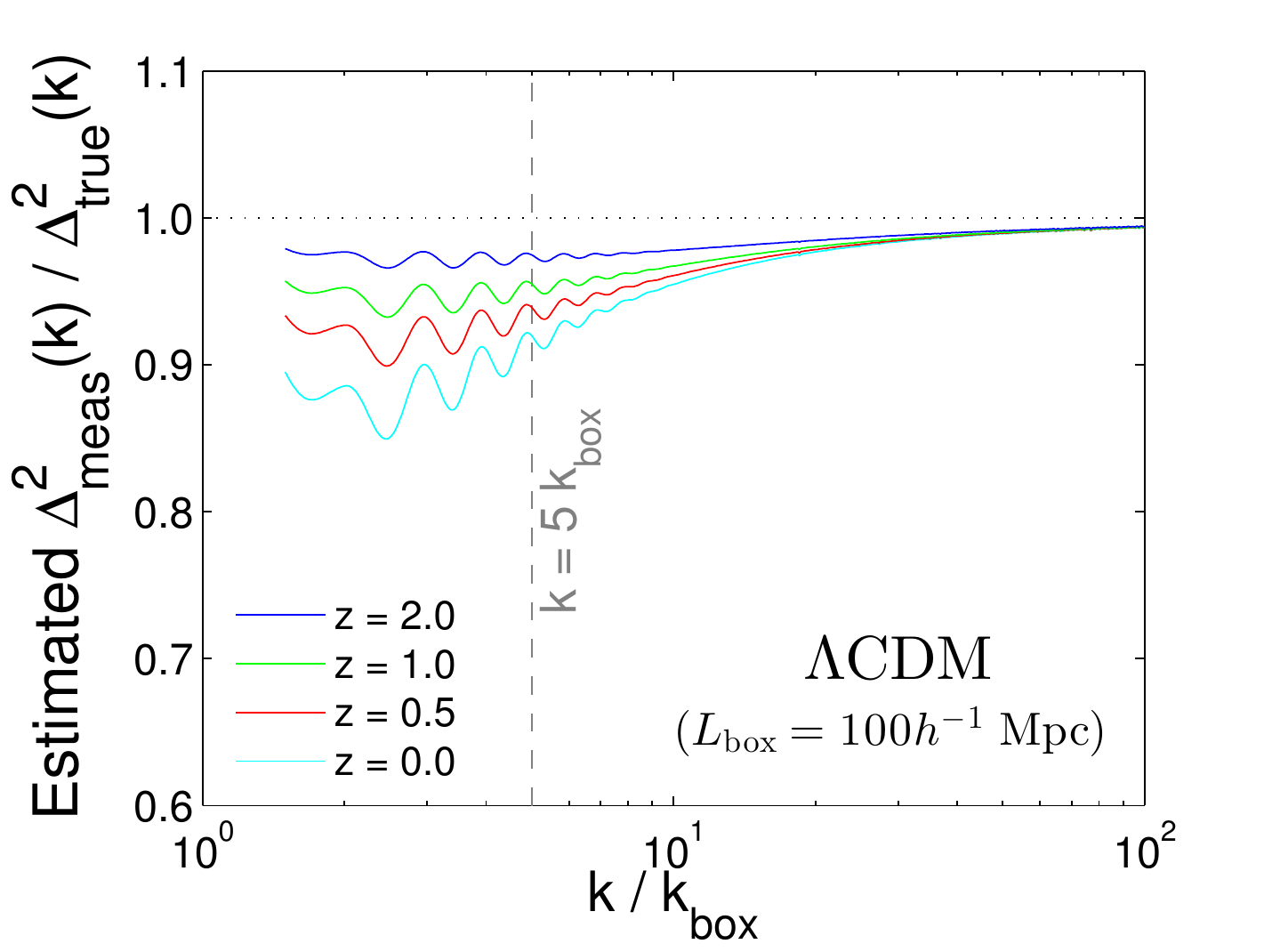, width=3.0in}\epsfig{file=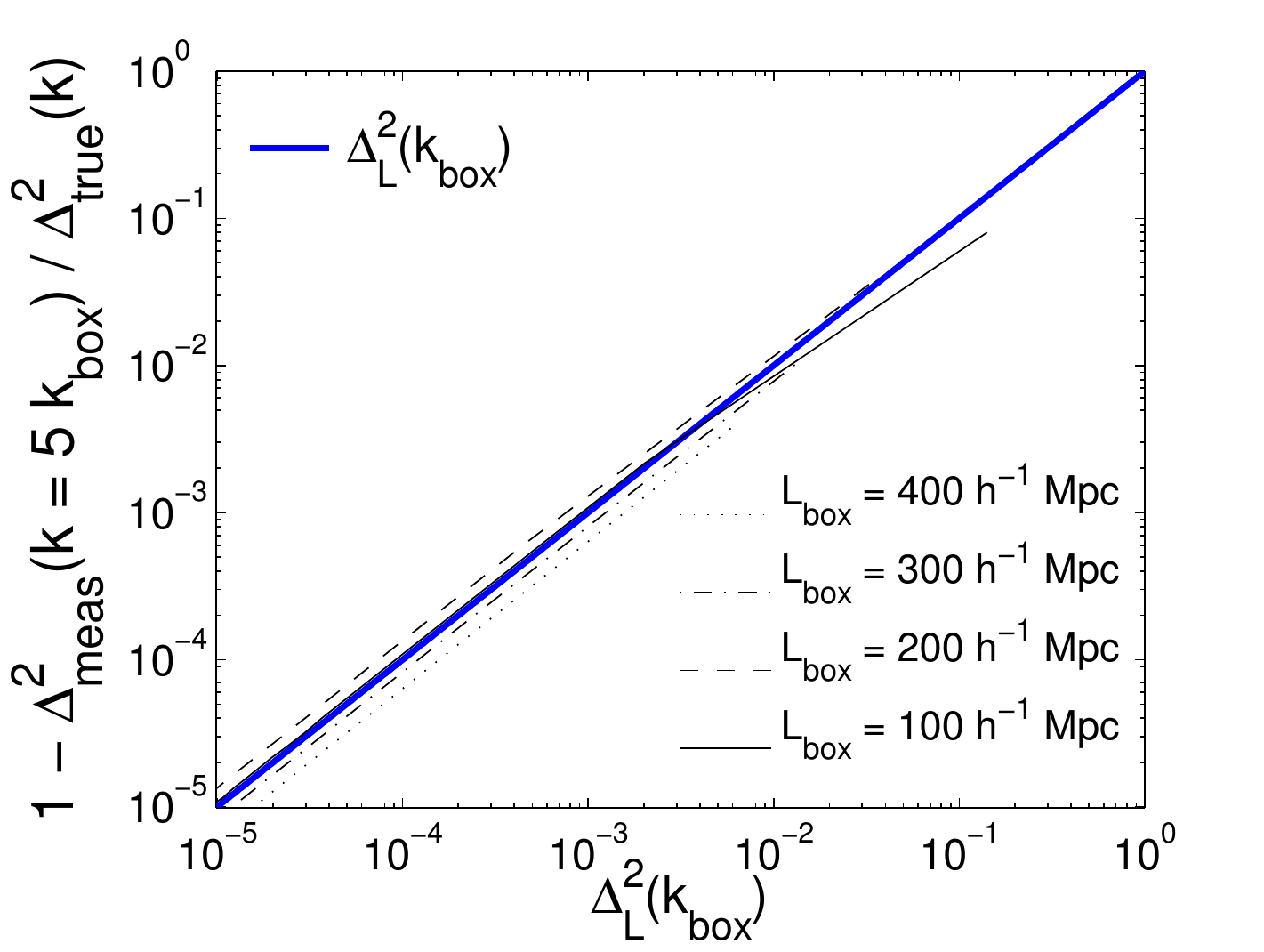,width=3.0in}}
\vspace{-0.4cm}
\caption{ Left panel: SPT estimated box-scale suppression of the mean power spectrum for $\Lambda$CDM simulations with $L_{\rm box} = 100 h^{-1}$~Mpc. Solid colored lines correspond to different redshifts as indicated in the legend. Right panel: Comparing the SPT estimated box-scale suppression of the power spectrum at $k = 5 \, k_{\rm box}$ for a variety of different redshifts and box sizes. The $x$-axis displays the results according to the level of dimensionless linear theory power on the scale of the box, $\Delta_L^2(k_{\rm box})$. A thick blue line shows the value of $\Delta_{\rm L}^2 (k_{\rm box})$ to show that, according to SPT, the non-linear power spectrum will be suppressed in proportion to $\Delta_{\rm L}^2 (k_{\rm box})$ at any redshift.}  \label{fig:lcdm}
\end{figure*}

In this section SPT predictions for the box-scale effects on the power spectrum in a $\Lambda$CDM cosmology will be considered ($\Omega_m = 0.25$, $\Omega_b = 0.0463$, $\sigma_8 = 0.8$, $h = 0.72$, $n_s = 0.97$). These parameters correspond to the fiducial cosmology used in \citet{Heitmann_etal2010} to perform various convergence tests for the non-linear power spectrum, including tests with varying the box size that are presented in their Fig.~6. By comparing the non-linear power spectrum measurements from their largest-box simulations ($L_{\rm box} = 2000 h^{-1}$Mpc, four realizations, $N = 1024^3$) to the same measurements from their smallest-box simulations ($L_{\rm box} = 234 \, h^{-1}$Mpc, 127 realizations, $N = 513^3$) the box-scale suppression of the non-linear power spectrum in the $L_{\rm box} = 234 \, h^{-1}$Mpc simulations can be robustly determined from their Fig.~6. This result can then be compared to the predicted box-scale suppression from SPT assuming a $\Lambda$CDM cosmology. 

Fig.~\ref{fig:coyote} presents these results, all coming from measurements or SPT estimates at $z = 0$. Because the suppression of the non-linear power spectrum in the $L_{\rm box} = 2000 h^{-1}$Mpc simulations is negligible for the scales in question, the non-linear power spectrum measurements for this box is referred to as $\Delta_{\rm true}^2(k)$ in Fig.~\ref{fig:coyote}. The measurements from the $L_{\rm box} = 234 h^{-1}$Mpc simulations are referred to as $\Delta_{\rm meas}^2(k)$. The ratio of $\Delta_{\rm meas}^2(k)$ and $\Delta_{\rm true}^2 (k)$ reveals the box scale suppression of the power spectrum\footnote{Since power spectra are typically evaluated at specific $k$-values relative to the box scale, the power spectrum for the $L_{\rm box} = 2000 \, h^{-1}$Mpc simulations needed to be interpolated to the $k$-values in the $L_{\rm box} = 234 \, h^{-1}$Mpc where the power spectrum was measured to construct this ratio. Also note that $k_{\rm box}$ in Fig.~\ref{fig:coyote} is equal to $2 \pi / L_{\rm box}$ where $L_{\rm box} = 234 h^{-1}$Mpc rather than $L_{\rm box} = 2000 h^{-1}$Mpc.}. While the measurement of the box-scale suppression is noisy, the $N$-body simulation results for $L_{\rm box} = 234 h^{-1}$Mpc fall a few percent below the ``true'' result for $k = 1-10 \, k_{\rm box}$\footnote{At higher $k$ the SPT estimate tends toward 1.0 (not shown) as in the powerlaw cases considered in Sec.~\ref{sec:results} and in the left panel of Fig.~\ref{fig:lcdm}.}. This was noticed earlier in \cite{Heitmann_etal2010} who comment that this few percent suppression of the power spectrum extends into the linear regime at the lowest $k$-values. It is satisfying that the perturbation theory prediction (gray line in Fig.~\ref{fig:coyote}) likewise predicts a few-percent suppression of the non-linear power spectrum for simulations with $L_{\rm box} = 234 h^{-1}$Mpc at $z = 0$.

To illustrate the redshift dependence of the box-scale suppression of the power spectrum, the left panel of Fig.~\ref{fig:lcdm} highlights a $\Lambda$CDM case with $L_{\rm box} = 100 h^{-1}$Mpc. The SPT estimated box-scale suppression is shown there, however there is no comparison to $N$-body simulations as in Fig.~\ref{fig:coyote}.  Overall SPT predicts of order 10\% suppression of the non-linear power spectrum by $z = 0$, which is significantly more suppression than the $L_{\rm box} = 234 h^{-1}$Mpc case considered in Fig.~\ref{fig:coyote}. 

To quantify the relationship between $\Delta_L^2(k_{\rm box})$ and the box-scale suppression of the power spectrum, the right-hand panel compares the predicted suppression of the power spectrum at $k = 5 \, k_{\rm box}$ to the value of $\Delta_L^2(k_{\rm box})$. This comparison is made for a variety of box sizes. Each line indicates the predicted suppression over a wide range of redshifts with $z = 0$ at the end of the line where the suppression is the largest and where $\Delta_L^2 (k_{\rm box})$ is the largest as well. Remarkably, this panel shows that independently of box size and redshift the suppression at $k = 5 \, k_{\rm box}$ is always within a factor of a few of $\Delta_L^2 (k_{\rm box})$ itself. Since $\Delta_L^2(k_{\rm box})$ is readily calculated from the box size and the linear theory power spectrum this result is a very useful rule-of-thumb for anticipating and avoiding box scale effects on power spectra from $N$-body simulations. For example, if the value of $\Delta_L^2(k_{\rm box})$ at redshift zero for some box size is, e.g., 0.03 = 3\% as it would be for $L_{\rm box} \approx 200 h^{-1}$ Mpc, then SPT predicts that the mean power spectrum at one fifth of the box scale will be suppressed by approximately this amount at redshift zero. At earlier redshifts it will be suppressed less than this. 

To be clear, this relationship between $\Delta_L^2 (k_{\rm box})$ and the box-scale suppression of the non-linear power spectrum comes entirely from SPT and these predictions should be further validated by $N$-body simulations. Fig.~\ref{fig:coyote} provides a crude validation of the SPT estimated suppression in $L_{\rm box} = 234 h^{-1}$ Mpc simulations at $z = 0$ from \cite{Heitmann_etal2010}. Validating the SPT predictions at higher precision and with a wider range of box sizes and redshifts (especially at the sub-percent-level precision investigated in the bottom left portions of the right panel of Fig.~\ref{fig:lcdm}) requires a careful, concerted effort that is beyond the scope of this paper.

\section{Summary and Conclusions}

Following ideas set forth by \citet{Seto1999}, the prospect of using standard perturbation theory (SPT) \cite{Vishniac1983} with a truncated linear power spectrum to estimate box-scale effects on power spectra from $N$-body simulations is here considered and compared with simulation results. Importantly, the truncated SPT calculation does not require knowledge of the random phases at the beginning of the simulation. Therefore the accuracy of $N$-body simulations can be estimated even when this information is unavailable or in the planning stage before the simulation is performed.

 This study presents simulations using power-law initial conditions with Fourier-space powerlaws of $n = -1.5$ and $n = -2$, which are similar to the effective slope of the $\Lambda$CDM power spectrum on the scales of BAO oscillations or smaller \cite[][Fig.~2]{Orban_Weinberg2011}. Powerlaw initial conditions are convenient for validating Seto's idea because the self-similar properties of these models allow the accuracy of simulation results to be judged without the need to run additional simulations.

The $n = -1.5$ and $n = -2$ simulation ensembles ($N = 512^3$, 25 realizations each) were run until the dimensionless power on the scale of the box, $\Delta_L ^2(k_{\rm box})$, reached the (extreme) value of 0.123. At the last output the $n = -1.5$ results showed a $\sim10$\% suppression of the non-linear power spectrum on scales near the box scale while the $n = -2$ results showed a suppression closer to $\sim 20$\%. The measured suppression from the simulation results were compared to the truncated SPT prediction versus $k / k_{\rm box}$, finding factor-of-two or better agreement between the truncated SPT model and the data. Though outside the scope of this study, the agreement could very likely be improved by using more sophisticated perturbation theory schemes. Note also that estimates for box-scale effects on other statistical quantities (e.g. the correlation function or halo mass function) can be obtained through their relation to the power spectrum.

Having investigated Seto's idea for estimating box-scale effects in power-law simulations, the $\Lambda$CDM case is considered. $N$-body simulation results presented in \citet{Heitmann_etal2010} were useful in confirming the accuracy of the SPT estimated box-scale suppression for $L_{\rm box} = 234 h^{-1}$Mpc simulations at $z = 0$. While much more precise comparisons to $N$-body simulations should be conducted, this result was encouraging for using SPT in this way. Other approaches in the literature for estimating the box scale suppression for the non-linear power spectrum were discussed and found to be ambiguous or less-than-quantitative.

SPT estimates for the redshift dependence of the box-scale suppression in $\Lambda$CDM simulations were also shown. SPT predicts that the power spectrum at $k = 5 \, k_{\rm box}$ (i.e. one fifth of the box scale) is suppressed by approximately the value of $\Delta_L^2 (k_{\rm box})$ regardless of the choice of $L_{\rm box}$ and redshift and over many orders of magnitude in $\Delta_L^2 (k_{\rm box})$. While this prediction needs to be confirmed in detail by $N$-body simulations, this result should be a broadly-applicable rule-of-thumb for quickly estimating the accuracy of power spectra measured from $\Lambda$CDM simulations. As an example, in $\Lambda$CDM simulations with $L_{\rm box} = 200 h^{-1}$~Mpc one expects $\sim3$\% suppression of the non-linear power spectrum near or around 40$h^{-1} $Mpc scales at redshift zero and less suppression at earlier redshifts. 

The early outputs from the power-law simulations presented here are the most precise and (both empirically and according to the truncated SPT estimate) the most accurate measurements of the self-similar evolution of $n = -1.5$ and $n = -2$ powerlaw initial conditions on quasi-linear scales ($k \lesssim k_{\rm nl}$) that exist in the literature. Appendix~\ref{ap:sf96} uses these results to examine the accuracy of the non-linear fitting functions for $n = -1.5$ provided by \cite{Orban_Weinberg2011} and for $n = -2$ provided by \cite{Widrow_etal09}. Appendix~\ref{ap:sf96} also confirms that Pajer \& Zaldarriaga's \cite{Pajer_Zaldarriaga2013} typo-corrected SPT prediction from \cite{Scoccimarro_Frieman1996,Scoccimarro1997} agrees better with simulations. Both of these findings proved useful in accurately inferring the box-scale suppression of the power spectrum in the powerlaw simulations. The power spectrum measurements from these powerlaw simulations are provided with this paper for future comparisons to more-sophisticated perturbation theory models. Along these lines, the author plans to perform detailed comparisons of these power spectrum measurements to the EFT model of Pajer \& Zaldarriaga \cite{Pajer_Zaldarriaga2013} in future work.

\section*{Acknowledgements}

Many thanks to the Ohio State University Center for Cosmology and
AstroParticle Physics for its support. I also thank Matias Zaldarriaga, 
Enrico Pajer and Katrin Heitmann for insightful correspondence.

\newpage

\appendix 
\onecolumngrid

\section{Estimating the Accuracy of both SPT and Previously-Published Non-Linear Fitting Functions \\ by Comparison to Simulation Outputs where the Suppression is Known to be Small}
\label{ap:sf96}

The inferred box-scale suppression of the non-linear power spectrum, $\Delta_{\rm meas}^2 (k) / \Delta_{\rm true}^2(k$), discussed in previous sections, relies on an accurate and precise knowledge of $\Delta_{\rm true}^2 (k)$. Rather than assume that this knowledge is perfect for the powerlaw models considered in Sec.~\ref{sec:results}, this appendix considers the accuracy of the non-linear fitting functions used to determine $\Delta_{\rm true}^2 (k)$. This task involves comparing the fitting functions to the $N$-body simulation results presented earlier but selecting only the early outputs when the clustering level on the scale of the box is relatively small, specifically $\Delta_{\rm L}^2 (k_{\rm box}) \leq 0.023$, and avoiding clustering scales that are too close to the box scale, i.e. rejecting measurements from $k \leq 5 \, k_{\rm box}$. These conservative choices enable the accuracy of the fitting functions to be very reliably measured. The data set is also quite useful for assessing the accuracy of the SPT prediction \emph{without} any truncation of the linear power spectrum. In earlier sections this is referred to as $\Delta_{\rm SPT}^2 (k)$. 

It should be emphasized that fitting functions for powerlaw simulations are always defined in terms of some kind of well-defined physical scale (e.g. the wavenumber $k$) divided by a relevant non-linear scale (e.g. $k_{\rm nl}$ where $\Delta_{\rm L}^2~(k_{\rm nl})~\equiv~1$). This is because the scale-free nature of the powerlaw initial conditions implies that the non-linear growth of perturbations, both in principle and in practice, should likewise evolve in a scale-free, a.k.a. ``self-similar'', way. This property of self-similarity dramatically simplifies the task of constructing a fitting function because each simulation output statistically resembles a scaled version of earlier and later outputs (c.f. \cite{Orban2011} Fig. 1.1 for an illustration). To illustrate this mathematically, using the non-linear dimensionless power spectrum, 
\begin{equation}
\Delta^2 (k) = \Delta_{\rm L}^2 (k) \Big[ 1 + f ( k / k_{\rm nl}) \Big] = \left( \frac{k}{k_{\rm nl}} \right)^{n+3}  \Big[ 1 + f ( k / k_{\rm nl}) \Big]  \label{eq:fit}
\end{equation}
where $f(k / k_{\rm nl})$ is the fitting function which will be different depending on the powerlaw $n$. The measured power spectrum from every simulation output should agree (within error bars) with this functional form with a suitable choice of $k_{\rm nl}$. Eq.~\ref{eq:fit} is the model used for $\Delta_{\rm true}^2 (k)$ in the powerlaw investigations in Sec~\ref{sec:results}. For $n = -1.5$, the fitting function comes from \citet{Orban_Weinberg2011}. For $n = -2$, the fitting function comes from \citet{Widrow_etal09}. This appendix compares these fitting functions to the powerlaw simulation data to determine how accurate they are.

\begin{figure*}[t]
\centerline{\epsfig{file=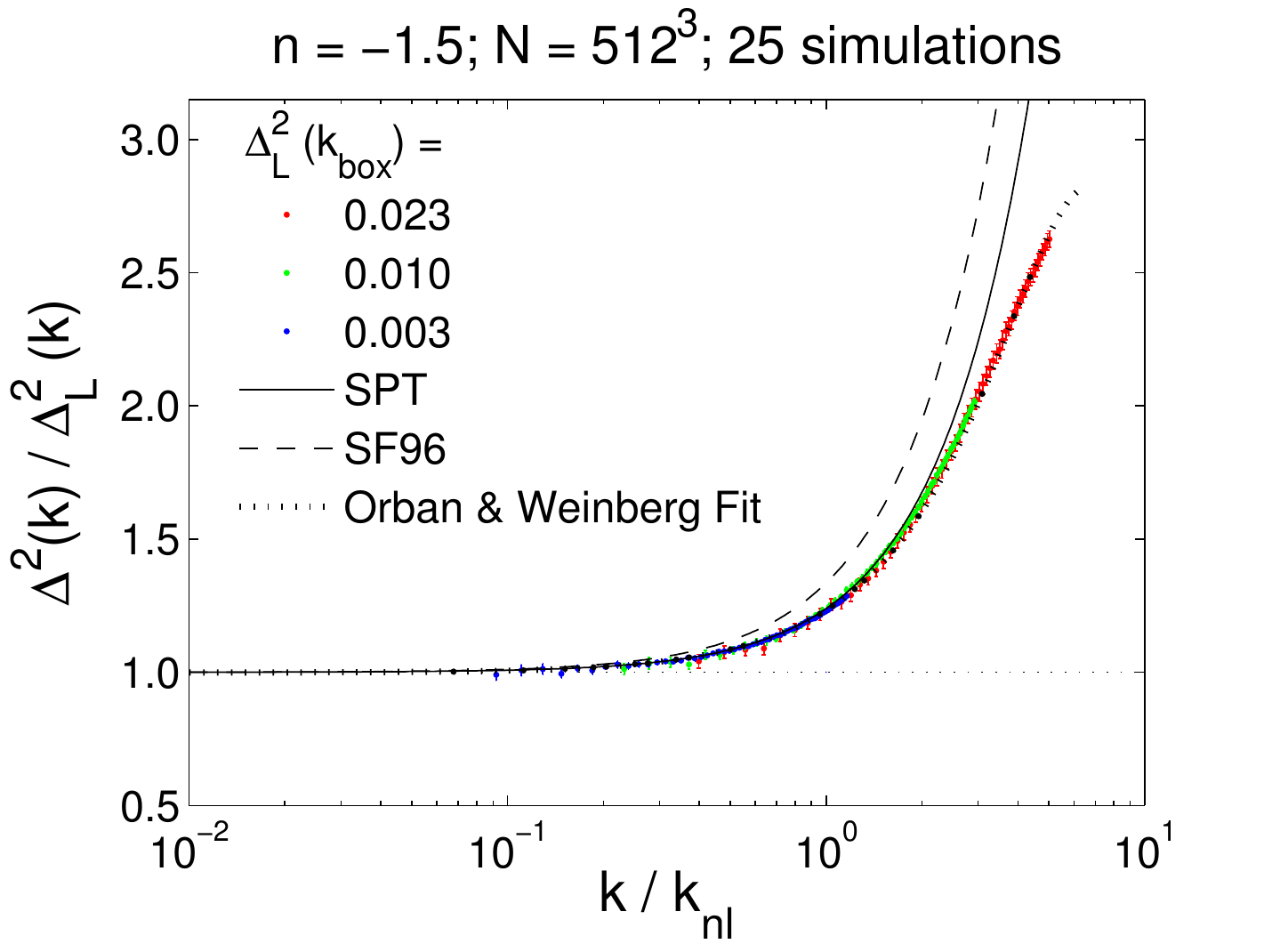,width=3.1in}\epsfig{file=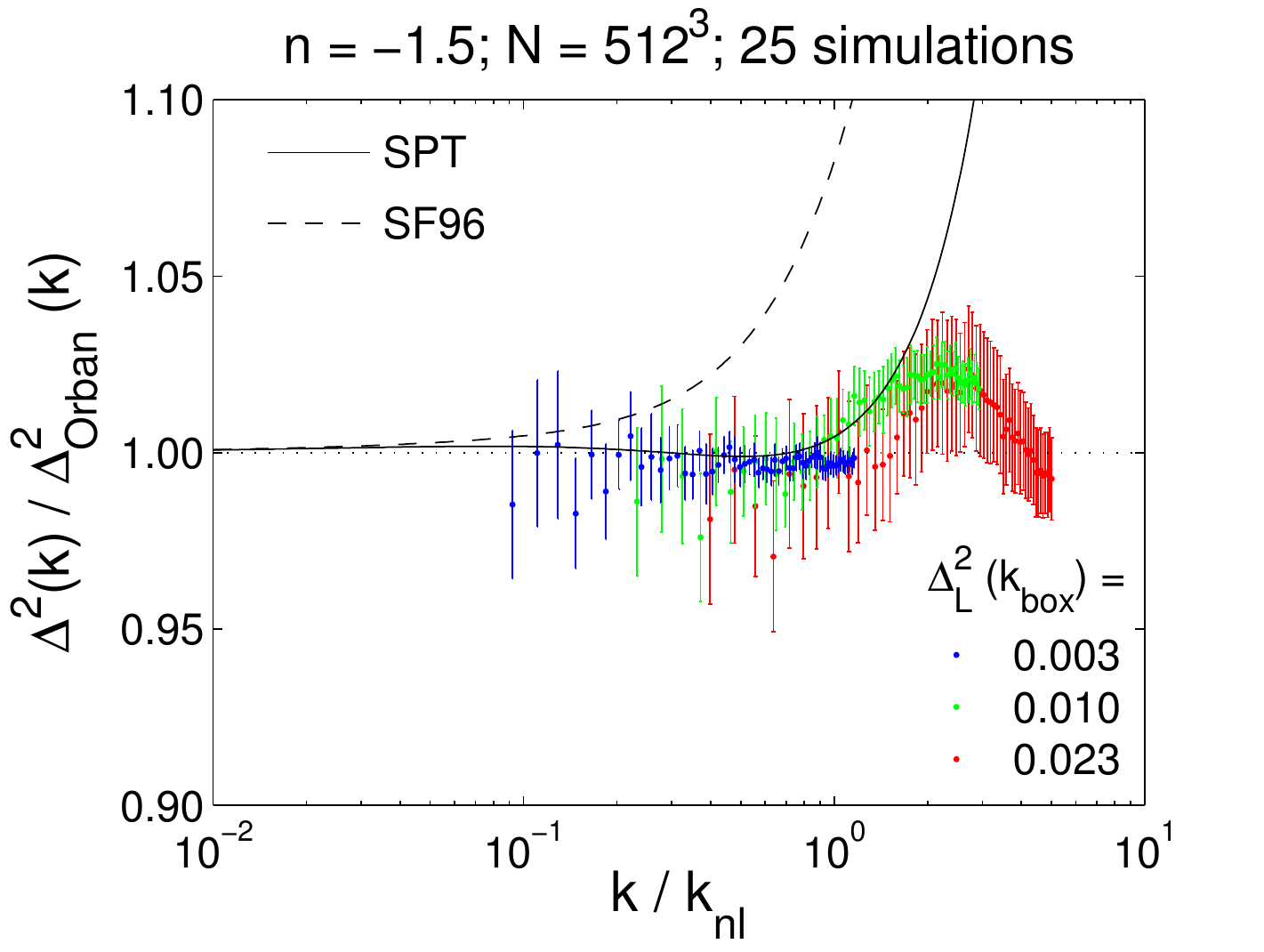,width=3.1in}}
\centerline{\epsfig{file=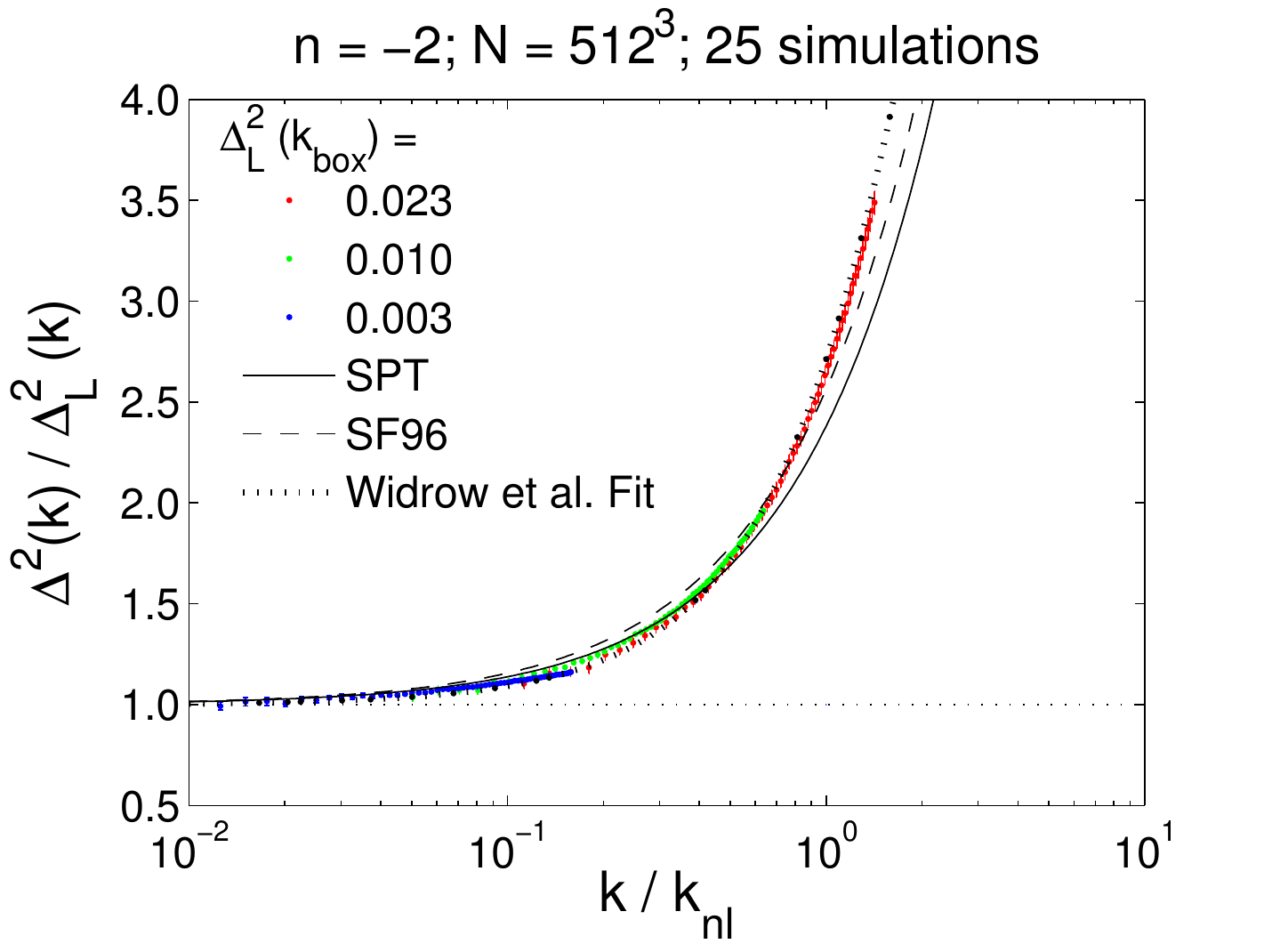,width=3.1in}\epsfig{file=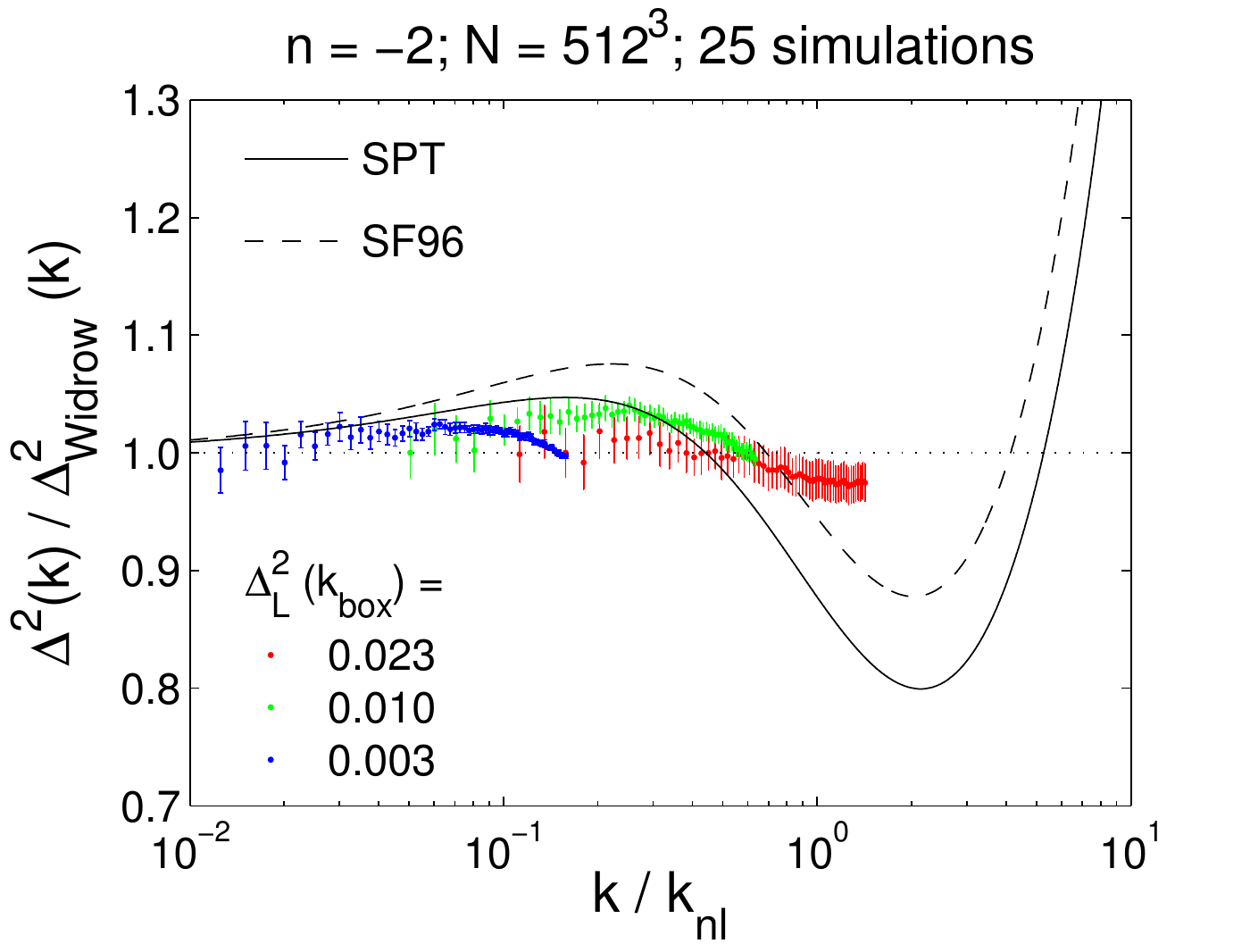,width=3.1in}}
\caption{
A comparison of simulation results (colored points with error bars) to perturbation theory models (black lines with different line types) for $n = -1.5$ (upper panels) and $n = -2$ (bottom panels). Unlike Fig.~\ref{fig:ptmiss}, the $x$-axis shows the wavenumber $k$ relative to the non-linear scale, $k_{\rm nl}$, where $k_{\rm nl}$ is defined by $\Delta_L^2 (k_{\rm nl}) \equiv 1$. Also shown with dotted black lines are previously-published fits to powerlaw simulations from either \citet{Widrow_etal09} or Orban \& Weinberg \cite{Orban_Weinberg2011}. In the right-hand panels the power spectrum results are shown relative to these non-linear fits. 
}\label{fig:sf96}
\end{figure*}

Fig.~\ref{fig:sf96} presents simulation data (colored points with error bars) from the power-law simulation ensembles discussed earlier in the text showing only the early few outputs and power spectrum measurements from wavenumbers $k > 5 \, k_{\rm box}$. The upper two panels show results from the $n = -1.5$ simulation ensemble presented earlier while the lower two panels show results from the $n = -2$ simulation ensemble. Note that unlike in previous figures the x-axes in Fig.~\ref{fig:sf96} show $k / k_{\rm nl}$ instead of $k / k_{\rm box}$. This choice highlights the self-similar shape of the non-linear power spectrum in the left-hand panels where the $y$-axes present $\Delta^2(k) / \Delta_L^2(k)$ where $\Delta_L^2(k)$ is the linear theory model for the power-law power spectrum.  The right-hand panels highlight the accuracy of the \citet{Orban_Weinberg2011} and \citet{Widrow_etal09} fitting functions.  To the extent that the measurements from each output lie along the same locus of points, this is strong evidence for the essential accuracy of the simulation ensemble\footnote{This is true for both the left and right panels because dividing by the \citet{Orban_Weinberg2011} and \citet{Widrow_etal09} fitting functions does not ``break'' self-similarity because these functions are defined in terms of $k / k_{\rm nl}$} in spite of any number of possible sources of error (e.g. box-scale effects, finite-particle effects, force resolution concerns, or subtleties in the generation of the initial conditions).  Historically, self-similar simulations were decisive in proving the accuracy of the first generation of cosmological $N$-body simulations \cite{Efstathiou_etal1988}. 

Regarding the right-hand column in Fig.~\ref{fig:sf96}, it should be reiterated that the simulation results presented there are significantly more precise for $k \lesssim k_{\rm nl}$ than previously-published non-linear fits from either \cite{Widrow_etal09} or \cite{Orban_Weinberg2011}. A close look at these plots indicate that the \citet{Widrow_etal09} fit is a few percent too low near $k \sim k_{\rm nl} / 10$, and that the Orban \& Weinberg fit is likewise a few percent too low at $k \sim 2 \, k_{\rm nl}$ but otherwise the agreement is typically within the measured 2$\sigma$ error bars. This mild tension is understandable given that the Orban et al. and Widrow et al. fits come from power spectrum measurements that have significantly larger error bars than the simulation results in Fig.~\ref{fig:sf96} because those studies performed many fewer simulations for each powerlaw than presented here. In principle, some differences may arise because of different choices regarding the initial conditions\footnote{Both studies use 2LPT \cite{Crocce_etal06} initial conditions and the Gadget2 code, however \citet{Widrow_etal09} begin their simulations at a less conservative starting point, $\Delta_{ic}^2 (k_{\rm Ny,p}) \approx 0.08$, than either the simulations presented here or in \cite{Orban_Weinberg2011}}, but since the fitting functions are constructed from multiple outputs and since these outputs exhibit self-similar behavior the impact of this and any number of other numerical details is greatly minimized.

The few-percent-or-better accuracy of the fitting functions was sufficient for the task of using these fitting functions to compare with simulation results at later epochs and a wide range of scales, including some scales where the box-scale suppression is significant. Thus, although the simulation results presented in Fig.~\ref{fig:sf96} are the most accurate and precise measurements of the $n = -1.5$ and $-2$ non-linear power spectra to date, constructing new fitting functions was not necessary. For the benefit of future studies the raw dimensionless power spectrum measurements including error bars are included with this paper. Making available these measurements \emph{with the error bars} will be very useful for those interested in checking the accuracy of perturbation theory models at high precision. 

Fig.~\ref{fig:sf96} also shows two expectations from perturbation theory. The SPT results in these plots (solid black lines) should be understood as the SPT predictions \emph{without} any truncation of the linear power spectrum at the scale of the simulation box (Eq.~\ref{eq:spt}, with $\alpha_{-1.5} = 0.239$ and $\alpha_{-2} = 1.38$ as reported in Pajer \& Zaldarriaga \cite{Pajer_Zaldarriaga2013}). In other words, it is the SPT prediction for the true non-linear power spectrum. 
Note that the SPT results differ from the SPT-based prediction from \citet{Scoccimarro_Frieman1996}, which is shown in Fig.~\ref{fig:sf96} with a dashed black line, because of a typo in one of their expressions that was identified by \cite{Pajer_Zaldarriaga2013}. The right-hand column in Fig.~\ref{fig:sf96} shows the simulation and perturbation theory predictions divided by the either the Orban \& Weinberg \cite{Orban_Weinberg2011} or \citet{Widrow_etal09} fit to previously-published simulation results. The results in Fig.~\ref{fig:sf96} show, for the first time, that correcting this problem significantly improves the agreement with the simulated non-linear power spectrum\footnote{The corrected SPT model is accurate enough that it can be used to create a highly-accurate, entirely (instead of partially) {\it ab initio} analytic expression for the non-linear power spectrum of the simplified BAO model in \cite{Orban_Weinberg2011}.}.

\bibliography{ms}
\bibliographystyle{apsrev}

\end{document}